\documentclass[a4paper]{jpconf}
\usepackage{graphicx}
\usepackage{amssymb}
\begin{document}
\title{Collapse models: from theoretical foundations to experimental verifications}

\author{Angelo Bassi}

\address{Department of Physics, University of Trieste, Strada Costiera 11, 34151 Trieste, Italy}

\address{Istituto Nazionale di Fisica Nucleare, Trieste Section, Via Valerio 2, 34127 Trieste, Italy}

\ead{bassi@ts.infn.it}

\author{Hendrik Ulbricht}

\address{School of Physics and Astronomy, University of Southampton, Southampton SO17 1BJ, United Kingdom}

\ead{h.ulbricht@soton.ac.uk}

\begin{abstract}

The basic strategy underlying models of spontaneous wave function collapse (collapse models) is to modify the Schr\"odinger equation by including nonlinear stochastic terms, which tend to localize wave functions in space in a dynamical manner. These terms have negligible effects on microscopic systems---therefore their quantum behaviour is practically preserved. On the other end, since the strength of these new terms scales with the mass of the system, they become dominant at the macroscopic level, making sure that wave functions of macro-objects are always well-localized in space. We will review these basic features. By changing the dynamics of quantum systems, collapse models make predictions, which are different from standard quantum mechanical predictions. Although they are difficult to detect, we discuss the most relevant scenarios, where such deviations can be observed.\end{abstract}

\section{Introduction}

There is a well-known and simple reason why quantum theory, in its standard formulation, is not satisfactory as a theory of nature. When applied to macroscopic systems, it predicts the existence of superpositions of different macroscopic states---Schr\"odinger's cat being dead + alive. However such superpositions do not match with the world as we perceive it. Hence the problem: The link between theory and physical phenomena (at least at the macroscopic level) is not clear~\cite{Bell}.

Over the decades, several ways to solve the problem has been presented. The textbook solution introduces the collapse of the wave function, which takes places at the end of each measurement process---essentially, when a microscopic systems interacts with a macroscopic object---and prevents the Schr\"odinger equation from turning microscopic superpositions into macroscopic ones. It is an easy practical way to dismiss the problem, but it cannot be taken too seriously, since it introduces an arbitrary and not better defined division between micro-macro, small-large. A fundamental theory of nature cannot be founded on such a vague and ill-defined basis.

Some advocate decoherence to solve the problem~\cite{Dec1,Dec2,Dec3}. Large objects like measuring devices, or cats, unavoidably interact with the surrounding environment, which cannot be fully controlled. Not having a full control on part of the universe the object interacts with, forbids in general the possibility of controlling and measuring quantum coherence for that object, hence no possibility of seeing Schr\"odinger's cat dead + alive. Again, this solution works out very well from the practical point of view, but not from the fundamental one. If the Schr\"odionger equation rules the universe, than the universe lives in a huge, complicated, highly nonclassical superposition, nothing that looks like the world as we know it~\cite{Nodec1,Nodec2}.

The situation improves slightly if combined with a many-worlds interpretation~\cite{MW1,MW2,MW3}. The huge, complicated, highly nonclassical superposition the universe lives in, actually refers not just to our universe, but to several ``parallel universes'', to use an expression taken from science fiction. Our universe corresponds to one term of the superposition. The emergence of probabilities, i.e. the statistics of experiments as we experience them, though, is still problematic.

In  recent years, the success of quantum  information technologies pushed part of the community to considering the  possibility that information itself is the foundation of physics~\cite{Info1, Info2, Info3}. It is a very fascinating idea, nevertheless it is hard to dismiss the question of J. Bell~\cite{Bell}: ``Information about what?", calling for a `reality', which information is about. And if this is a case, than the fundamental theory should be about that reality, first of all.

There are two other alternatives, which provide a simple and rather straightforward explanation of quantum phenomena: Bohmian mechanics~\cite{Bohm1,Bohm2,Bohm3} and collapse models~\cite{CM0,CM1,CM2,CM3,CM4,CM5}. The first theory tells that the world is made of (point-like) particles and gives the equation for the motion of the particles. Everything follows from that. The second theory tells that the world is made of wave-packets, and gives the equation for the motion of these wave-packets. Everything follows from that. At least at the non-relativistic level. Here we show how this is accomplished, in the case of collapse models.

\section{How to modify the Schr\"odinger equation}
According to the Ghirardi-Rimini-Weber (GRW)and Continuous Spontaneous Localization (CSL) models~\cite{CM1,CM2}, perhaps the two most popular collapse models, the Schr\"odinger equation is modified by adding nonlinear and stochastic terms, which induce the collapse of the wave function in space. The modification is such that these models are compatible with all known experimental data about quantum phenomena, and at the same time they make sure that macroscopic systems are never in funny macrocopic superpositions---Schr\"odinger's cat is always either dead or alive.

\begin{figure}[t]
\includegraphics[height=0.6\textwidth,width=1\textwidth]{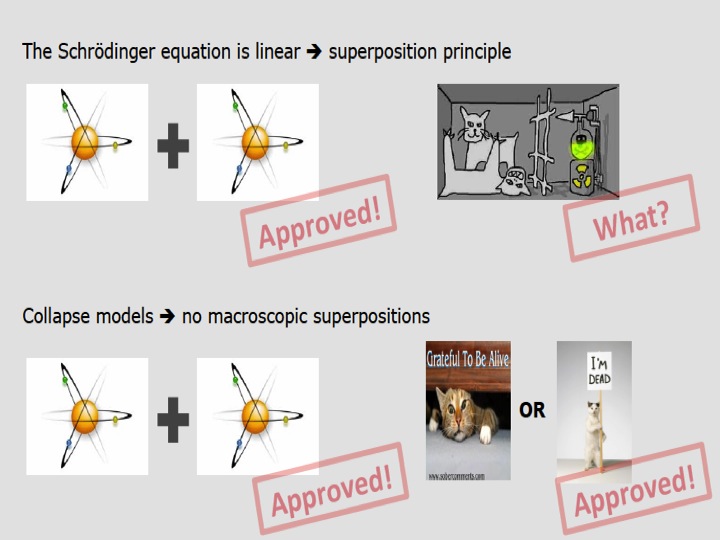}
\caption{\label{label} According to standard quantum theory, the superposition principle holds at all scales. Therefore Schr\"odinger's cat can be in a dead + alive state, which does not correspond to any classical state of it. According to collapse models, the superposition is (almost) right only at the microscopic scale, while it is more and more violated when moving toward larger and larger systems, to the point that macroscopic systems like Schr\"odinger's cat is always in a classical state, either dead or alive.}
\end{figure}

In order to better understand the role played by collapse models as an alternative to standard quantum theory, let us take a broader perspective. Let us suppose for a moment that these models were not already on the market, and let us ask the question: How can we modify the Schr\"odinger equation in a consistent way, without violating basic principles of physics? This issue was considered repeatedly in the literature, and was given a precise formulation by S. Adler~\cite{AdlerTD}: The only possible modifications to the Schr\"odinger equation, which preserve the norm of the state vector and do not allow for superluminal signaling, are only those corresponding to collapse models.  More recently, also S. Weinberg~\cite{Weinberg} raised the same question and has given the renewed interest in the subject.

The (more or less) most general form of the collapse equation is:

\begin{eqnarray} \label{eq:sdgr}
{d} \psi_t & =  & \left[ -i H {d}t + \sum_{k=1}^{n} \left(  L_k - \ell_{k,t} \right){d} W_{k,t} \right.  - \left. \frac{1}{2} \sum_{k=1}^{n} \left(L^{\dagger}_k L^{\phantom{\dagger}}_k  - 2 \ell_{k,t}L_k + |\ell_{k,t}|^2\right){d}t\right] \psi_t, \quad \\
& & \nonumber \\
\ell_{k,t} & \equiv & \displaystyle \frac12\langle\psi_t,(L^{\dagger}_k + L^{\phantom{\dagger}}_k)\psi_t\rangle.
\end{eqnarray}
The first term on the right-hand-side represents the usual quantum evolution (unitary part). The other terms are responsible for the collapse of the wave function: the wave function is driven towards one of the common eigenstates of the operators $L_k$ (assuming that they exist, as it is the case for all collapse models considered in the literature), with a probability equal to the Born rule~\cite{CM2,Hugh}. Note that these terms are both nonlinear (the quantum average $\langle \psi | ... | \psi \rangle$ enters the equation) and stochastic ($W_{k,t}$ are standard independent Wiener processes), as it should be in order to have a collapse process.

\section{The CSL model}

In the above equation, the only relevant degree of freedom is given by the form of the collapse operators $L_k$. As stated here above, the nonlinear terms tend to collapse the wave function towards one of the common eigenstates of $L_k$, therefore choosing these operators amount to choosing the preferred basis for the collapse.

If we wish to collapse the wave function in the energy basis, we choose $L_k = H$. If we wish to collapse the wave function in momentum, than we pick the momentum operator. But since the main duty of collapse models is to make sure that macroscopic objects are always localized in space, the most direct---and in many ways most natural---choice is to pick the position operator, or some function of it. This is what happens in the GRW model, the CSL model, and all other most relevant collapse models. 

Here we report the equation for the CSL model:
\begin{equation}
d |\psi_t \rangle = \left[ - \frac{i}{\hbar} H dt + \sqrt{\lambda} \int d^3x ( N({\bf x}) - \langle N({\bf x}) \rangle_t) dW_t({\bf x}) - \frac{\lambda}{2} \int d^3x ( N({\bf x}) - \langle N({\bf x}) \rangle_t)^2 dt \right] |\psi_t\rangle,
\end{equation}
where $H$ is the standard quantum Hamiltonian, in the QFT language, $N({\bf x}) = a^{\dagger}({\bf x}) a({\bf x}) $ is the particle density operator, $\lambda$ the coupling constant which sets the strength of the collapse mechanism, and $W_t({\bf x})$ a family of independent standard Wiener processes with zero average and correlation:
\begin{equation}
\mathbb{E} [W_t({\bf x})W_s({\bf y})] = \delta(t-s)e^{-(\alpha/4)({\bf x-y})^2}.
\end{equation}

The choice of the particle {\it density} as the collapse operators corresponds precisely to the desire of having wave functions localized in {\it space}, i.e. it is a good alternative to the position operator in the formalism of QFT.

\section{The collapse mechanism. Choice of the parameters.}

The CSL mode is defined in terms of two parameters: the collapse strength $\lambda$ and the spatial correlation function of the noise $r_C = 1/\sqrt{\alpha}$. The first parameter sets the strength of the collapse mechanism for one particle, and through the {\it amplification mechanism} which we will now explain, also for a macroscopic system. The second parameter tells which kind of superpositions are effectively suppressed, and which not. The situation is pictorially explained in Fig.~2.

\begin{figure}[t]
\includegraphics[width=1\textwidth]{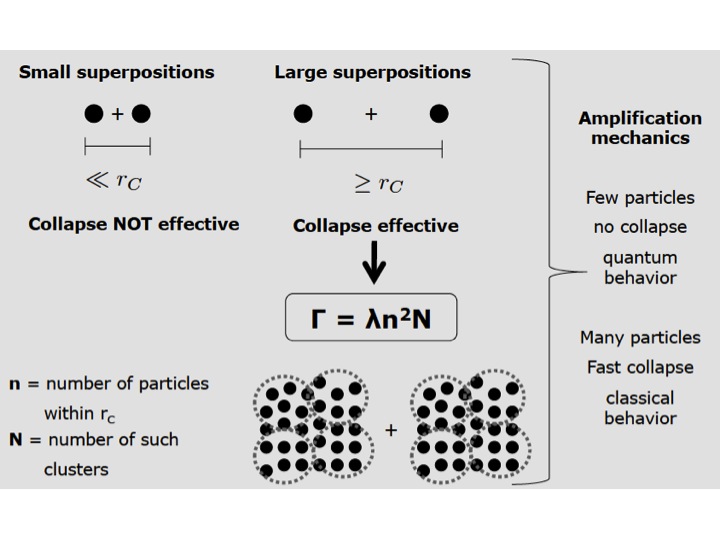}
\caption{\label{label}The {\it amplification mechanism} is the crucial property of collapse models: the collapse rate for the center of mass of a composite object is a function of the size and number of constituents of the system. The larger the system, the stronger the collapse rate. This is how collapse models can accommodate, within one single dynamical equation, both the quantum properties of microscopic systems and the classical properties of macroscopic objects.}
\end{figure}

Let us consider a single particle (a nucleon) in a superposition of two localized states, whose relative distance in space is $d$. If $d$ is smaller than $r_C$, then the noise---so to say---does not see such a superposition as a `macroscopic' superposition (the two states are too close to each other), and does not modify it in a significant way. On the other hand, if $d$ is much larger than $r_C$, then the noises sees it as a macroscopic superposition and tends to localized it to one of the two states. The rate is given by $\lambda$. Therefore $r_C$ sets the threshold between microscopic and macroscopic superpositions.

Let us now consider a composite system, an object, again in a superposition of two different states, distant relative to each other more than $r_C$. The collapse mechanism becomes difficult to unfold analytically, due to the complexity of the dynamics for the system. Nevertheless one can work out a easy approximative rule to estimate the effect~\cite{Adlerlb}. The center of mass of the system collapses with a rate:
\begin{equation}
\Gamma = \lambda n^2 N,
\end{equation}
where $\lambda$ is the collapse rate of a single constituent (a nucleon), $n$ is the number of constituents contained in a volume of radius $r_C$ (therefore it is a measure of the density of the system) and $N$ counts how many such volumes can be accommodated in the system (therefore a measure of the total volume of the system). 

The important message is that the collapse rate of a composite system scales with its size: the bigger the system, the faster the collapse. This is precisely the reason why collapse models can accommodate, within a single dynamical equation, both the quantum properties of microscopic systems (few constituents) and the classical properties of macroscopic objects (many constituents). In doing this, collapse model explain the quantum-to-classical transition, how the probabilistic and wavy nature of atoms and molecules gives rise to the world of classical physics as we experience it, when they glue together to form larger and larger objects. 

As a marginal note, one can see that the collapse rate scales partly linearly and partly quadratically with the number of constituents. The quadratic increase is due to the identity of particles: if the noise tries to collapse a particle, it has to collapse all other particles identical to it, in order to preserve the symmetry of the wave function. This quadratic increase is not present in the original GRW model, which deals only with distinguishable particles.

Given this, the relevant question is: which values can one reasonably assign to $\lambda$ and $r_C$? The question is not easy to answer from the theoretical point of view, as there are only very rough guidelines one can follow. As we explained here above, $r_C$ sets the division between micro and macro distances. A reasonable value to take, first suggested by GRW~\cite{CM1}, is:
\begin{equation}
r_C \simeq 10^{-7} \textrm{m},
\end{equation}
which is much bigger than typical inter-atomic distances in matter ($\sim 10^{-10}$m), but much smaller than human-size distances ($\sim 0.1$mm). Of course, this value could fluctuate a few orders of magnitude, but this is the best one can say at the moment.

There is much less of a consensus regarding the value $\lambda$ should have. In particular, two very different values were proposed, The first one was suggested by GRW:
\begin{equation}
\lambda_{\textrm{\tiny GRW}} = 10^{-16} \textrm{s$^{-1}$},
\end{equation}
which sets the quantum-to-classical (i.e. no-collapse vs. collapse) transition at the border between the mesoscopic and macroscopic world. This mean that particles, atom, small and large molecules and even those objects which are invisible at the naked eye, but otherwise are rather complex, all behave in a fully quantum mechanical way. Only when we reach the level of objects of our daily-life experience (dust grains, tables, chairs) the collapse becomes effective and protects wave functions from being in macroscopic superpositions. Therefore, what proposed by GRW is more or less the weakest, most conservative value for $\lambda$ one can take, which guarantees that the macroscopic world behaves in a classical way.

In more recent years, Adler suggested to take a much stronger value~\cite{Adlerlb}:
\begin{equation}
\lambda_{\textrm{\tiny Adler}} = 10^{-8} \textrm{s$^{-1}$}.
\end{equation}
The motivation is that the collapse process should be effective in all those situations which we would refer to as a measurement process, like the process of latent image formation in photography. A simple estimate of the number of atoms involved in this process shows  that, in order for the collapse to be effective within the reaction time which leads to the formation of the latent image,  one needs to take $\lambda$ some eight orders of magnitude stronger than what suggested by GRW. In this way, the threshold between quantum and classical is moved to the border between the microscopic and mesoscopic world.

This huge difference between the two proposals might be puzzling, but merely reflects our ignorance about the validity of the superposition principle in a vast domain. We know it holds for elementary particles, up to large molecules. We know that macroscopic systems are never observed in macroscopic superpositions. What happens in between, no one knows. It is a matter of intense theoretical and most of all experimental debate, as we will now discuss.

\section{Remarks on experimental tests}
While there is a broad range of possible experimental tests for collapse models, we are here focusing on only some of those: matter-wave interferometry experiments where a macroscopic quantum superposition state has to be generated, recently proposed high-resolution spectroscopy tests in the frequency domain and tests based on a tiny violation of energy conservation: the Pearle disc proposal. We shortly sumarize the state of the art of realised experiments and how close they are to actually test some of the proposed collapse models. We also emphasise the timeline for the experimental realisation of those tests, which have not yet been realised.  A brief summary of the most prominent experimental tests can be found here~\cite{BassiAdler}.  

\subsection{Quantitative measures to classify experiments}
Clearly a set of parameters has to be specified to decide if a specific experiment is able to test the superposition principle in the macroscopic domain and therefore collapse models. Such parameters are the mass of the particle, the spatial size of the single-particle superposition state and the time duration of its existence. A recently introduced measure, called Macroscopicity $\mu$ which includes all three parameters is essential to objectively compare experiments~\cite{NimmrichterMacro}.  

For the experiment it means the superposition state has to be associated to a large mass, large spatial separation between the positions where the particle is delocalised and a long (life)time for the existence of this spatial superposition states. It becomes clear that not every matter-wave experiment would be sensitive to test Macroscpicity as defined above. For instance atomic matter-waves, while having an impressive momentum spread in modern experiments~\cite{Kasevich}, the mass is too light for a test. Also the formation of loosely bound ensembles of many identical atoms (Bose Einstein condensate) cannot be of help. Contrary,  the very massive mechanical cantilevers, which are unwed for quantum opto-mechanic experiments~\cite{aspel} and have been shown that can be prepared in the vibrational ground state, possess typically a too small spatial separation to be really macroscopic [see table in reference~\cite{NimmrichterMacro}]. But it would not be wise to have a last word about this today.  Experimental development is vital and due to ever ongoing inventions, which could put back such systems on the map. For instance a way further for mechanical cantilevers would be to transfer the quantum superposition state on two identical cantilevers as proposed here~\cite{Xuereb}. 

\subsection{Experiments using matter waves}
The interest in testing the limits of validity of quantum mechanics has increased in the last decades. The test of the break of the linearity of the 
Schr\"odinger  equation, the lost of unitarity are all different names for the same thing, namely to test if quantum mechanics continues to be the correct theory even for macroscopic systems and if the quantum to classical transition can be explained by a universal mass-scaling mechanism. Clearly, while many competing theories and theoretical concepts try to explain this transition two of the  main contenders are decoherence theory and the here discussed collapse models. There is an increasing number of experimental groups around the world, rushing towards creating quantum superpositions, which are as macroscopic as possible. In the nutshell, a very massive single particle superposition, with a large spatial and temporal extension has to be generated to test the quantum superposition principle. Here we want to look into some conceptual ideas to experimentally test quantum linearity. 

\subsection{Test as of today}
So far the answer is affirmative, but the systems employed are still rather small. The Macroscopicity record is hold by the Arndt group in Vienna by molecule interferometry~\cite{molrev} of organic molecules of mass above 10,000 amu (atomic mass units)~\cite{Gerlich2011, Eibernberger}. The experiment is a so called Talbot-Lau interferometer, where the diffraction grating is a structure made of a retroreflected laser beam, based on the Kaptiza-Dirac effect~\cite{Gerlich2007, Hornberger2009}.  The natural extension of those particle beam experiments is proposed here for heavy particle and cluster beams. The OTIMA interferometer has been already realised~\cite{Haslingertheo, Haslinger}. Now intense heavy-particle gas-phase beams have to be realised~\cite{Juffmann}. Macroscopicity and how this experiments can test collapse models is worked out here~\cite{NimmrichterCSL}.

\subsection{New proposals for tests using matterwave interferometry}
That experiments have motivated others to come up with new proposals for macroscopic matter-wave experiments. Others propose the use of levitated particles (optical or magnetic) to generate a spatial quantum superposition state of an even larger (more massive) particle~\cite{Bose, romeroopto, romeromag}. Ideas range from performing fast tests while the particle stays always in the trap~\cite{Bosespin, Juan} to drop particles after initial state preparation~\cite{romeroopto, James}.   Recent progress in the manipulations and preparation of the centre of mass motion allows for the cooling to about 10 mK, which is sufficient for one of the proposals to generate a superposition state~\cite{James}. However a number of experimental challenges (mainly related to build a nanoparticle source in vacuum) remain before the experiments can be realised.  

In the first case the stability of the trap has to be enormous and limits the actual lifetime of the superposition state to typically less than 10 $\mu$s. In some schemes internal (spin) states are prepared by a sequence of magnetic pulses in a resonance spectroscopic Ramsey technique and a closed interferometer has to be achieved within the time constraints given by dephasing and decoherence of the spin states. This internal state interferometry is coupled to the centre of mass motion of the magnetic particle and therefore represents a matter-wave interferometer.   

The free fall / drop experiments are limited by the acceleration of the particle in Earth's gravitational field. For a particle of increasing mass the de Broglie wavelength of the accelerated particle by $g$ gets so small that matter-wave experiments are almost impossible. The limit on mass is at the one million amu level. Such experiments will be able to test the CSL value with Adler's value for $\lambda$~\cite{NimmrichterCSL, CM5}, but other models like the original GRW can not be approached and need a larger mass for the superposition.
Therefore, MAQRO  has been proposed to explore possibilities for quantum superposition experiments with nanoparticles in space~\cite{Maqro}. Related conceptual and practical aspects have to been worked out~\cite{Rainernjp}.  

A further alternative is the transfer of superposition from matter-wave beam to a pair of mechanical cantilevers. A new concept for generating a almost macroscopic quantum superposition state of a large mass has be proposed by us [22]. The basic idea is that an atomic matter-wave is imprinted on a pair of identical mechanical cantilevers. State of the art quantum optics calculations show that the superposition state can survive this transfer, which if done in a real experiment would boost test in a new regime ($10^{13}$ amu). The status of this project requires now a careful design of the experiment and major funding to set it up. This experiment, if possible to realised, can fast all proposed collapse models~\cite{NimmrichterCSL, CM5}.

\section{Experiments in the frequency domain}
The general strategy for the aforementioned matter-wave interferometry tests is to create superpositions in space, and test them with interferometers. Basically all groups follow this strategy. Recently we suggested an alternative route: to test quantum mechanics not in the time domain (see whether macroscopic superspositions exist and persist in time), but in the frequency domain~\cite{Bahramihigh, Bahrami}. The idea is that deviations from quantum theory modify the spectral properties of atoms and molecules, and therefore can be checked with sophisticated spectroscopic techniques.  The analysis shows that high-resolution experiments in the not too far future should be possible. For more details, see~\cite{Bahrami}. Currently we are investigating this general effect applied specifically on opto-mechanical systems. The goal is to understand if parameter settings in today's opto-mechanis experiments can test collapse models.

\section{Other experiment}
A-normal (collapse induced) Brownian motion: Another interesting idea, which is worth revisiting is an effect first described by Philipp Pearle~\cite{Pearle}: A micrometer-sized disc is (frictionless) mounted inside an ultra-high vacuum chamber at very low temperatures to avoid heating and collisions of the disc. Then, collapse models predict, that due to stochastic noise, the disc should start to rotate and a quantitative effect is predicted and compared to classical Brownian motion caused by particle collision and the quantum version from e.g. laser shot noise. The parameter regime is practically accessible but various details of the concrete experimental setup have to be explored in a feasibility study.  

\section{Outlook}

Collapse models are a viable alternative to quantum mechanics, avoiding the interpretational problems of the standard formulation of the theory. They are a well founded theory, since they consists of just a dynamical equation for the wave function, and everything else follows from that. This is what theories should be like. In particular, it can  be proven~\cite{Salv} that the whole quantum formalism of measurements (observables as self-adjoint operators, the role of eigenvalues and eigenvectors, the Born rule ...) can be deduced, not pustulated, from the dynamical equation of collapse models, when applied those specific physical situations, which we call measurements. 

On the other hand, it is somehow clear that collapse models are phenomenological, and call for an underlying theory out of which the quantum world---with the collapse---emerges, perhaps in a similar why to how thermodynamics emerges from classical statistical mechanics. The difficult question is to say something about this possible underlying theory. The only framework we are aware of, is the one provided by Adler with Trace Dynamics~\cite{AdlerTD}. The work is still in progress, but already represents a promising way of understanding the quantum world as emerging from a deeper level of reality. 

Another interesting question is if the collapse is related to gravity. This idea has been on the market for a long time, and perhaps the strongest input has been given by Penrose~\cite{Penr1,Penr2}. There are good reasons to explore this possibility: gravity is a nonlinear theory, and nonlinearity is precisely what one needs to kill superpositions. Moreover, gravity becomes stronger, the larger the systems, just like the collapse process. Models have been developed~\cite{DP,sn2,sn2}, but a breakthrough is still needed, in order to make it convincing that the collapse can be really linked to gravitational effects.

Also experimentally we are experiencing and exciting boost in possibilities and ideas on how to test collapse models. We predict that in the next three years some of the CSL models, such as the one proposed by Adler, will have been experimentally tested by matter-wave experiments on Earth while  other experiments will be pushed forward to test even more macroscopic quantum states. If a state collapse is observed, specific scaling of loss of quantum features of the superposition state (such as visibility of the quantum interference contrast) with experimental controllable parameters (such as internal and external temperature of the particle in superposition or the number of other particles in its environment)  can be used to distinguish the collapse mechanism.  

\section*{Acknowledgments}
AB thanks Gerhard Gr\"ossing for organizing the conference EmQM13 on Emergent Quantum Mechanic, and  the Fetzer Franklin fund for its financial support. He acknowledges financial support from the EU project NANOQUESTFIT and INFN.  He also thanks the COST Action MP1006 ``Fundamental Problems in Quantum Physics". HU thanks the UK funding agency EPSRC (EP/J014664/1). Both authors thank the Foundational Questions Institute (FQXi) and the John F Templeton Foundation for financial support.


\section*{References}
\begin{thebibliography}{9}

\bibitem{Bell} J.S. Bell, {\it Speakable and Unspeakable in Quantum Mechanics}, Cambridge University Press (1988).

\bibitem{Dec1} E. Joos, H. D. Zeh, C. Kiefer, D. Giulini, J. Kupsch and I.O. Stamatescu, {\it Decoherence and the Appearance of a Classical World in Quantum Theory}, Springer-Verlag (2003).

\bibitem{Dec2} H.-P. Breuer and F. Petruccione, {\it The Theory of Open Quantum Systems}, Oxford University Press (2002)

\bibitem{Dec3} M.A. Schlosshauer {\it Decoherence and the Quantum-to-Classical Transition}, Springer  (2008). 

\bibitem{Nodec1} A. Bassi and G. C. Ghirardi, Phys. Lett. {\bf A 275}, 373 (2000).

\bibitem{Nodec2} S.L. Adler, Stud. Hist. Philos. Mod. Phys. {\bf 34}, 135 (2003).

\bibitem{MW1} H. Everett, {\it Theory of the Universal Wave function}, Thesis, Princeton University, (1956, 1973).

\bibitem{MW2} H. Everett,  Rev. Mod. Phys. {\bf 29}, 462 (1957). 

\bibitem{MW3} B.S. DeWitt, Phys. Today, {\bf 23}, 30 (1970). 

\bibitem{Info1} G. M. D�Ariano, in: ``Philosophy of Quantum Information and Entanglement'', edited by A. Bokulich and G. Jaeger, Cambridge University Press, Cambridge (2010).

\bibitem{Info2} G. Chiribella, G. M. D'Ariano, P. Perinotti,  Phys. Rev. A {\bf 84} 012311 (2011).

\bibitem{Info3} \v{C}aslav Brukner,  Physics {\bf 4}, 55 (2011).

\bibitem{Bohm1} P.H. Holland, {\it The Quantum Theory of Motion: An Account of the de Broglie-Bohm Causal Interpretation of Quantum Mechanics}, Cambridge University Press (1995).

\bibitem{Bohm2} D. D\"urr and S. Teufel, {\it Bohmian Mechanics: The Physics and Mathematics of Quantum Theory}, Springer (2009)

\bibitem{Bohm3} D. D\"urr, S. Goldstein and N. Zangh\`i, {\it  Quantum Physics Without Quantum Philosophy}, Springer (2013).

\bibitem{CM0} P. Pearle, Phys. Rev. D {\bf 13}, 857 (1976).

\bibitem{CM1} G.C. Ghirardi, A. Rimini and T. Weber, Phys. Rev. D {\bf 34}, 470 (1986).

\bibitem{CM2} G.C. Ghirardi, P. Pearle and A. Rimini, Phys. Rev. A {\bf 42}, 78 (1990).

\bibitem{CM3} A. Bassi and G.C. Ghirardi, Phys. Rept. {\bf 379}, 257 (2003).

\bibitem{CM4} S.L. Adler and A. Bassi, Science {\bf 325}, 275 (2009).

\bibitem{CM5} A. Bassi, K. Lochan, S. Satin, T.P. Singh and H. Ulbricht, Rev. Mod. Phys. {\bf 85}, 471 (2013). 

\bibitem{AdlerTD} S.L. Adler, {\it Quantum Theory as an Emergent Phenomenon: The Statistical Mechanics of Matrix Models as the Precursor of Quantum Field Theory}, Cambridge University Press (2009).

\bibitem{Weinberg} S. Weinberg, Phys. Rev. A {\bf 85}, 062116 (2012).

\bibitem{Gisin} N. Gisin, Hel. Phys. Acta {\bf 62}, 363 (1989).

\bibitem{Lin} G. Lindblad, Commun. Path. Phys. {\bf 48}, 119 (1976).

\bibitem{GKS} V. Gorini, A. Kossakowski and E.C.G. Sudarshan, J. Math. Phys. {\bf 17}, 821 (1976).

\bibitem{Hugh} S. L. Adler, D. C. Brody, T. A. Brun, and L. P. Hughston,  J. Phys. A: Math. Gen. {\bf  34}, 8795 (2001).

\bibitem{Adlerlb} S.L. Adler, Journ. Phys. A {\bf 40}, 2935 (2007).

\bibitem{Salv} A. Bassi, G.C. Ghirardi and D.G.M. Salvetti,  Journ. Phys. A {\bf 40}, 13755 (2007).

\bibitem{Penr1} R. Penrose, Gen. Rel. Grav. {\bf 28}, 581 (1996).

\bibitem{Penr2} L. Diosi, Phys. Lett. A {\bf 120}, 377 (1987).

\bibitem{DP} L. Diosi, Braz. J. Phys. {\bf 35}, 260 (2005).

\bibitem{sn1} L. Diosi, Phys. Lett. A {\bf 105}, 199 (1984).

\bibitem{sn2} D. Giulini and A. Grossardt.  Class. Quant. Grav. {\bf 28}, 195026 (2011).

\bibitem{BassiAdler} A. Bassi and S. Adler, Science 325, 275 (2009).

\bibitem{NimmrichterMacro} S. Nimmrichter and K Hornberger, Physical Review Letters 110, 160403 (2013).

\bibitem{NimmrichterCSL}, S. Nimmrichter, K. Hornberger, P. Haslinger, and M. Arndt, Phys. Rev. A 83, 043621 (2011).

\bibitem{aspel} M. Aspelmeyer, T. J.  Kippenberg, and F. Marquardt, arXive:1303.0733 (2013).

\bibitem{molrev} K. Hornberger, S. Gerlich, P. Haslinger, S. Nimmrichter, and M. Arndt,  Rev. Mod. Phys. 84, 157 (2012).

\bibitem{Kasevich} S.-W. Chiow, T. Kovachy, H.-C. Chien, and M. A. Kasevich Phys. Rev. Lett. 107, 130403 (2011).

\bibitem{Xuereb} A. Xuereb, H. Ulbricht, and M. Paternostro, Scien. Rep. 3, 3378 (2013).

\bibitem{Gerlich2011} S. Gerlich, S. Eibenberger, M. Tomandl, S. Nimmrichter, K. Hornberger, P. J. Fagan, J. T�xen, M. Mayor, and M. Arndt, Nat. Comm. 2, 263(2011).

\bibitem{Eibernberger} S. Eibenberger, S. Gerlich, M. Arndt, M. Mayor and J. T�xen,  Phys.Chem. Chem. Phys. 15, 14696 (2013).

\bibitem{Gerlich2007} S. Gerlich, L. Hackerm�ller, K. Hornberger, A. Stibor, H. Ulbricht, F. Goldfarb, T. Savas, M. M�ri, M. Mayor and M. Arndt, Nat. Phys. 3, 711 (2007). 

\bibitem{Hornberger2009} K. Hornberger, S. Gerlich, H. Ulbricht, L. Hackerm�ller, S. Nimmrichter, I. V. Goldt, O. Boltalina and M. Arndt, New J. Phys. 11  043032 (2009).

\bibitem{Haslingertheo} S. Nimmrichter, P. Haslinger, K. Hornberger and M. Arndt, New J. Phys. 13, 075002 (2011).

\bibitem{Haslinger} P. Haslinger, N. D�rre, P. Geyer, J. Rodewald, S. Nimmrichter, and M. Arndt, Nat. Phys. 9, 144�148 (2013).

\bibitem{Juffmann} T. Juffmann, H. Ulbricht, and M. Arndt,  Rep. Prog. Phys. 76 086402 (2013).

\bibitem{Bose} D. Home and S. Bose, Phys. Lett. A 217, 209 (1996).

\bibitem{romeroopto} O. Romero-Isart,  A. C. Pflanzer, F. Blaser, R. Kaltenbaek, N. Kiesel, M. Aspelmeyer, and J. I. Cirac, Phys. Rev. Lett. 107, 020405 (2011).

\bibitem{romeromag} O. Romero-Isart, L. Clemente, C. Navau, A. Sanchez, and J. I. Cirac, Phys. Rev. Lett. 109, 147205 (2012).

\bibitem{Bosespin}  M. Scala, M. S. Kim, G. W. Morley, P. F. Barker, and S. Bose, Phys. Rev. Lett. 111, 180403 (2013).

\bibitem{Juan}  Z.-Q. Yin, T. Li, X. Zhang, L. M. Duan, arXiv:1305.1701 (2013).

\bibitem{James} J. Bateman, S. Nimmrichter, K. Hornberger, and H. Ulbricht,  arXive:1312.0500 (2013).

\bibitem{Maqro} R. Kaltenbaek, G. Hechenblaikner, N. Kiesel, O. Romero-Isart, K. C. Schwab, U. Johann, M. Aspelmeyer, Exp. Astr. 34 pp 123 (2012).

\bibitem{Rainernjp} G. Hechenblaikner, F. Hufgard, J. Burkhardt, N. Kiesel, U. Johann, M. Aspelmeyer, and R. Kaltenbaek, arXiv:1309.3234 (2013).

\bibitem{Bahramihigh} M. Bahrami, S. Donadi, L. Ferialdi, A. Bassi, C. Curceanu, A. Di Domenico, B. C. Hiesmayr, Scie. Rep. 3, 1952 (2013).

\bibitem{Bahrami} M. Bahrami, A. Bassi, and H. Ulbricht, arXive:1309.5889 (2013).

\bibitem{Pearle} B. Collett, and P. Pearle, Found. Phys. 33, 1495 (2003).


\end{thebibliography}
\end{document}